\documentclass{aa}
\usepackage{graphicx}
\usepackage{natbib}
\bibpunct{(}{)}{;}{a}{}{,}
\begin{document}

\titlerunning{Microlensing constraints on the Galactic Bulge
Initial Mass Function}
\title{Microlensing constraints on the Galactic Bulge\\ Initial Mass Function}

\authorrunning{Calchi Novati et al.}
\author{S.~Calchi Novati\inst{1,2} 
\and F.~De Luca\inst{3}
\and Ph.~Jetzer\inst{3} 
\and L.~Mancini \inst{1,2} 
\and G.~Scarpetta\inst{1,2}
}

\institute{
%1
Dipartimento di Fisica ``E. R. Caianiello'', 
Universit\`a di Salerno, Via S. Allende, 84081 Baronissi (SA), Italy \and
%2
Istituto Nazionale di Fisica Nucleare, sezione di Napoli, Italy\and
%3
Institute for Theoretical Physics,  University of Z\"urich, 
Winterthurerstrasse 190, 8057 Z\"urich, Switzerland 
}
\date{Received 7 August 2007/ Accepted 16 November 2007}

\abstract{\\
\emph{Aims.} We seek to probe the Galactic bulge IMF
starting from microlensing observations.\\
\emph{Methods.} We analyse the recent results of the microlensing
campaigns carried out towards the Galactic bulge presented
by the  EROS, MACHO and OGLE collaborations. In particular,
we study the duration distribution of the events. 
We assume a power law initial mass function, $\xi(\mu)\propto \mu^{-\alpha}$,
and we study the slope $\alpha$ both in the brown dwarf 
and in the main sequence ranges.
Moreover, we compare the observed and expected optical depth profiles.
\\
\emph{Results.} 
The values of the mass function slopes are strongly 
driven by the observed timescales of the microlensing events. 
The analysis of the MACHO data set gives, for the main sequence stars, 
$\alpha=1.7 \pm 0.5$, 
compatible with the result we obtain with the EROS and OGLE data sets,
and a similar,  though less constrained slope for brown dwarfs.
The lack of short duration events in both  EROS and OGLE data sets, on the other hand,
only allows the determination of an \emph{upper} limit in this range of masses,
making the overall result less robust.
The optical depth analysis gives a very good agreement
between the observed and the expected values,
and we show that the available data do not allow one to discriminate
between different bulge models.
\keywords{Gravitational lensing - Galaxy: Bulge - Galaxy: stellar contents}}
\maketitle

%----------------------------------
\section{Introduction} \label{sec:intro}
%----------------------------------

Gravitational microlensing is an established
tool for the study and the characterisation
of faint compact objects located between the observer
and the source stars. It was originally proposed
as a tool for the detection of dark matter
in the form of MACHOs \citep{pacz86}. Searches
towards the Magellanic Clouds by the MACHO \citep{macho00}
and the EROS groups \citep{eros07} have placed strong constraints on the 
possible contribution of a MACHO population to the dark matter halo (for a discussion
see, e.g., \citealt{mancini04,novati06}). A few results have also been obtained
with observational campaigns towards M31
by the POINT-AGAPE \citep{novati05} and the MEGA \citep{mega06} collaborations.
On the other hand, the Galactic bulge soon proved
to be an almost as interesting target,
if not more \citep{pacz91,kiraga94} and
indeed, by now, the number of observed microlensing events
along this line of sight is by two orders of magnitude
larger than those observed towards the Magellanic Clouds and M31.
In this case, any contribution from a dark matter MACHO population is expected
to be extremely small compared to that of either bulge or disc stars \citep{griest91b}, 
so that these studies in principle allow us
to constrain the inner Galactic structure.
In particular, microlensing observations in this direction
have been very important for the assessment of the Galactic 
triaxial, bar-like, structure \citep{pacz94,zhao95,zhao96a,zhao96b,bissantz97,gyuk99}.

Recently, the MACHO \citep{popowski05}, OGLE  \citep{sumi06} and 
EROS \citep{hamadache06} collaborations presented the results
out of their several-year campaigns towards the Galactic bulge.
A remarkable result is the agreement among the different collaborations
for the \emph{optical depth}, in accord 
with theoretical expectations \citep{evans02,bissantz02,hangould03}.

While the determination of the optical depth allows the study
of the density distribution of the lenses,
a more detailed analysis of
the shape of the microlensing lightcurves carries much
information on the parameters of the lenses.
Of particular relevance is the possibility
of studying the \emph{mass spectrum} of the lenses.
This approach  is based on the relationship
between the observed event duration, the Einstein time,
and the mass of the lens $t_\mathrm{E} \propto \sqrt{\mu_l}$.
Even if the exact analytical formula
shows a dependence also on other unknown physical
parameters such as the distances of both lens and source,
the relative velocity between them as well
as the configuration of the particular lens event,
a few conclusions are made possible
by the rather large set of observed events at our disposal
together with a few reasonable assumptions
on the space and velocity distributions of both lenses and sources.
A key step to reach the aforementioned
agreement of the optical depth between theory
and experiments has been the acknowledgment
of the severe \emph{blending} problem resulting in
the choice of restricting the sample of source
stars to the \emph{red clump giant} subset \citep{popowski01}.
In turn, this is essential in the framework
of a mass spectrum analysis because of the bias introduced in the evaluation
of the Einstein time for blended events. 

The determination of the mass function
using the results of microlensing searches has
been addressed by several authors.
\cite{hangould96} consider a sample of MACHO and OGLE
events. Through a likelihood analysis they determine
the slope of a power law mass function to be 2.1 in the mass
range $(0.04-10)~\mathrm{M}_\odot$. 
\cite{jetzer94b} and \cite{grenacher99} use the mass moments method
to place constraints on the lens masses. 
In particular, starting from a sample of 41 MACHO events and assuming a Salpeter
profile in the mass range $(1-10)~\mathrm{M}_\odot$, \cite{grenacher99}
constrain the mass function minimum mass and 
slope below $1~\mathrm{M}_\odot$, finding
$0.012~\mathrm{M}_\odot$ and $2.0$ respectively.
Overall, therefore, there is an agreement in attributing a rather
large fraction of events to the brown dwarf lens population.
On the other hand, \cite{peale98} finds no compelling
evidence for such a contribution, and evaluates
the slope for a power law mass function in the mass range
$(0.08-2)~\mathrm{M}_\odot$ to be in the range $2.2-2.5$.
All of these analyses, we recall, used the complete sample
of detected events, not restricted to those with red clump giant sources.
\cite{bissantz04}, considering only red clump giant sources, 
find a good agreement with the MACHO observed timescale using 
a mass function with a large contribution from the brown
dwarf population (with a power law slope 2.35 in the mass range
$(0.04-0.35)~\mathrm{M}_\odot$). \cite{woodmao05}
extend the \cite{zoccali00} slope 1.3 down to a minimum
mass of 0.$03~\mathrm{M}_\odot$ compared with the
OGLE observed timescale.

In the present paper our aim is to make use
of the most recent observational results towards
the Galactic bulge to study 
the mass spectrum of the \emph{bulge} lens population.
The structure of the paper is as follows.
Sect.~\ref{sec:models} is devoted to the description of the models
we use. In Sect.~\ref{sec:anaml} we point out
a few particular features of the usual microlensing quantities
upon which we base our method of analysis.
In Sect.~\ref{sec:res} we present and discuss our main results. 
In Sect.~\ref{sec:end} we conclude.

%----------------------------------
\section{Models} \label{sec:models}
%----------------------------------

In this section we introduce, describe the features, and fix the parameters 
of our ``fiducial'' model
for the bulge and the disc needed to evaluate the microlensing quantities
that we use in the analysis. Furthermore,
we discuss a series of changes in the more critical parameters
that we use to test the robustness of our results.

%----------------------------------
\subsection{Density distributions} \label{sec:density}
%----------------------------------

%----------------------------------
\subsubsection{The bulge} \label{sec:bulge}
%----------------------------------

It is now acknowledged that the Galactic bulge
has a box-like (tri-axial) structure.
In an analysis of clump giant stars,
\cite{stanek97} explored several analytical distributions to describe the bulge. 
As a fiducial model we use their
model E2, which gives the best agreement 
with the observational data, where $\rho(r)=\rho_0 \exp(-r)$,
with\footnote{Here, as in Sect.~\ref{sec:vel} where
we discuss the velocity distribution,
the coordinates $x,y,z$ indicate the principal axes
of the component considered, namely, either of the bulge or of the disc.} 
$r=\sqrt{(x/x_0)^2+(y/y_0)^2+(z/z_0)^2}$,
$x_0=890~\mathrm{pc}$ and axis ratio values $x_0:y_0:z_0=10:4.3:2.8$,
an inclination angle of the bulge major axis with respect
to the line of sight of $\alpha=23.8^\circ$
(the bulge is oriented with its longer axis pointing towards us for
positive longitude values).
More recently \cite{rattenbury07a} carried out a similar
analysis with a much larger sample of stars.
As a result, the model E2 is again favoured, with axis ratio values
suggesting a more prolate structure, $x_0:y_0:z_0=10:3.5:2.6$,
and a more restricted range of bulge inclination values is given,
$\alpha\sim (24^\circ-27^\circ)$. 
In the analysis we keep using the \cite{stanek97} values, 
and we test our results against those of \cite{rattenbury07a}. 
We truncate the bulge at a corotation
radius $R_\mathrm{C}=3.5~\mathrm{kpc}$ \citep{bissantz02}.
The bulge inclination angle value 
is still the subject of a somewhat lively debate.
Values in the range $\alpha\sim (10^\circ-30^\circ)$
have been given by several authors, together with different values for the
axis ratio (e.g. \citealt{dwek95,sevenster99,picaud04}),
but recently also much larger values have been suggested.
Indeed, \cite{cabrera07} (and reference therein) 
discuss a more complicated inner Galactic structure 
with the co-existence of a double structure, composed
of a long ($\sim 4~\mathrm{kpc}$) thin and lighter bar located 
at low Galactic latitudes, $|b|<2^\circ$, and out to high Galactic longitude,
with an extremely large value for the inclination angle $\sim 43^\circ$,  and a distinct 
triaxial bulge with smaller inclination angle, $\sim 13^\circ$.
The star count results of GLIMPSE \citep{glimpse05}, in the $l=10^\circ-30^\circ$ range,
seem to confirm this result. 
Such a structure may of course give rise to interesting
microlensing signatures, however the currently available data
are not suitable for its study, as 
they exclude the Galactic plane region 
and are mostly restricted to events
observed at small Galactic longitudes.
(The EROS collaboration \citep{eros07}
evaluates a bulge orientation angle of $49^\circ\pm 8^\circ$
even if they observe only the region out to $|l|\sim~10^\circ$ 
and do not observe any field for $|b|<1^\circ$
and only a very few in the  band out to $|b|\sim~2^\circ$.)
The issue of the bulge inclination has also
been discussed in \cite{wood07} in the framework
of an analysis of the microlensing optical depth.
As an alternative bulge distribution, \cite{hangould95,hangould03} use 
the model G2, favoured by an analysis of the \emph{COBE} DIRBE observations \citep{dwek95},
that they correct for small ($r<700~\mathrm{pc}$) galactocentric distance
with the \cite{kent92} model. 
We compare these two models to the observed optical depth.
The total bulge mass is usually evaluated in the range
$M_\mathrm{bulge}\sim (1-2)~10^{10}\mathrm{M}_\odot$ \citep{blum95,zhao96a,dehnen98}.
Lacking any compelling constraints we choose to normalise the bulge distribution
to the observed value of the microlensing optical depth (Sect.~\ref{sec:tau}).
Throughout the paper we use $R_0~=~8~\mathrm{kpc}$ as the
value for the distance to the Galactic centre.

%----------------------------------
\subsubsection{The disc} \label{sec:disc}
%----------------------------------
The \emph{profile} of the disc distribution is better constrained than that of the bulge,
although the value of the parameters that characterise it
is subject to debate. In order to parametrise the model 
we follow closely \cite{hangould03}
who use a $\mathrm{sech}^2$ (exponential) profile for the thin (thick) components
and normalise the distribution so as to obtain a local stellar 
density of $\Sigma_0=36~\mathrm{M}_\odot~\mathrm{pc}^{-2}$.
We note however that \cite{hangould03} attribute 
a rather large density fraction to the ``thick'' disc,
whereas this component is usually reported to contribute only to a minor
fraction of the overall density (e.g. \citealt{dehnen98,vallenari06}).
For our fiducial model we assume, as compared to that of \cite{hangould03},
the extreme case where we set to zero the thick disc contribution
(in their notation, we use $\beta=0$ instead of $\beta=0.565$);
moreover, we fix the value of the local disc density
in agreement with the normalisation of the disc mass function (Sect.~\ref{sec:imf}).
We have then tested our results using the values of \cite{hangould03} and also
the \cite{freudenreich98} profile characterised by a decreasing density
towards the Galactic centre.
As already pointed out by \cite{hangould03}, we find that
our results do not depend significantly upon the disc model
as the bulge component gives by far the dominant
contribution to the observed events.

%----------------------------------
\subsection{Kinematic models} \label{sec:vel}
%----------------------------------
To evaluate the microlensing rate we have to specify
the velocities of the components involved.
For bulge and disc stars we take into account
both the bulk and random components of motion.
For the former,
for both disc and bulge we assume a solid body rotation out 
to $R_\mathrm{cut}$ and at outer radii a flat rotation with
$R_\mathrm{cut}=2~(1)~\mathrm{kpc}$ 
and $V_\mathrm{max}=220~(50)~\mathrm{km/s}$ for
the disc (bulge) component respectively.
As the bulge value is less well constrained
\citep{blum95,dehnen00,bissantz03,minchev07,rich07},
we have tested our results varying the bulge component by 30\% 
to larger and smaller values.
Furthermore, we take into account the solar motion.

For the random component, we assume the
velocity distributions to follow an anisotropic Gaussian profile.
For the disc dispersion we use
$\sigma_x=20~\mathrm{km/s}$ and
we consider a linear increase towards the
Galactic centre for the remaining components
with $(\sigma_y,\sigma_z)=(30,20)~\mathrm{km/s}$
and $(\sigma_y,\sigma_z)=(75,50)~\mathrm{km/s}$
in the local neighbourhood and at the Galactic centre 
respectively \citep{hangould95}.
For the bulge, whose velocity dispersions are
not as well constrained,
we consider two somewhat opposite cases.
As a first approach, we follow \cite{hangould95}
and fix the dispersion values using the virial theorem
as applied to the bulge distribution \citep{blum95}.
For our fiducial model we obtain $\sigma_{x,y,z}=(112.5,86.1,72.1)~\mathrm{km/s}$.
In Sect.~\ref{sec:imf2} we investigate the effects 
of changes with respect to our fiducial model.
Whenever we modify either the central density or the pattern speed of the bulge
component, we adjust
the dispersion values of the bulge according to the prescription
of the virial theorem. In only one case we arrive at rather significant
differences (beyond a few percent), namely, when we consider as
a disc model that of \cite{hangould03}. Indeed, in that case,
our evaluation of the bulge central density decreases by about 25\% (Sect.~\ref{sec:tau}),
implying $\sigma_{x,y,z}=(96.3,74.3,62.8)~\mathrm{km/s}$. 

As a second estimate, we make use of recent observational
results \citep{kozlowski06,rattenbury07b} and use
$(\sigma_l,\sigma_b)=3.0,2.5\,\mathrm{mas\,yr}^{-1}$,
with $\sigma_\mathrm{los}\sim 110~\mathrm{km/s}$ \citep{BM98}.
For a bulge inclination of $\alpha=23.8^\circ$ and $R_0=8~\mathrm{kpc}$
we then evaluate the dispersion along the bulge principal axes to be
$(109.4,114.8,94.8)~\mathrm{km/s}$.

%----------------------------------
\subsection{Mass functions} \label{sec:imf}
%----------------------------------
The main aim of the present work is to analyse the
microlensing events to place constraints
on the mass function of the bulge stars. 
\cite{zoccali00} study the bulge mass function in the range $(0.15-1)~\mathrm{M}_\odot$,
finding a good fit to the data with a IMF power law, 
$\xi(\mu)\propto \mu^{-\alpha}$, with $\alpha=1.3\pm0.1$. 
They also propose a power law
with a change of slope at $0.5~\mathrm{M}_\odot$ and
$\alpha \sim 1.4,\,2.0$ respectively below and above
this threshold. Overall, this result is compatible
with the previous analysis of \cite{holtzman98}.
The more difficult part of the mass spectrum
to be explored is the low mass tail, including very low mass main sequence stars
and the brown dwarf range. In their analysis of microlensing
events towards the bulge, \cite{hangould03} extend
the \cite{zoccali00} mass function down to well below the
hydrogen mass burning limit, at $0.03~\mathrm{M}_\odot$,
and the same was done more recently by \cite{woodmao05}.
\cite{gould00} describes 
how to treat remnants, assuming that all of the stars
with mass above 1~M$_\odot$ have by now entered the remnant phase.
Given a slope of the IMF in this mass range, \cite{gould00}
proposes $\alpha=2$, it is then possible to evaluate
the number and mass fractions due to each of these components
(white dwarfs, neutron stars and black holes).

Following the previous results, we assume
a power law mass function for both the brown dwarf and
the main sequence ranges. We introduce two parameters,
the slopes $\alpha_\mathrm{BD},\,\alpha_\mathrm{MS}$
in the mass ranges $(0.01-0.08)~\mathrm{M}_\odot$, 
$(0.08-1.0)~\mathrm{M}_\odot$
respectively, that we want to constrain. 
According to analyses carried out
for the disc, the slope should change below
the hydrogen burning limit (e.g. \citealt{kroupa07}).
We also test the effects on our results
of two changes on the bulge mass function,
namely  we introduce a slope change at 0.5~M$_\odot$,
using $\alpha=2$ above this limit \citep{zoccali00}, and we move the 
lower brown dwarf limit from 0.01 to 0.04~M$_\odot$.
We follow \cite{gould00} to deal with the remnants,
with $(0.6, 1.35, 5.0)~\mathrm{M}_\odot$ taken as the mass values for
white dwarfs, neutron stars and black holes respectively. 
Besides the value $\alpha_\mathrm{rem}=2$,
in this mass range we will further test our result with the higher value
$\alpha_\mathrm{rem}=2.7$, as suggested by disc results. Note that for every
pair of values $(\alpha_\mathrm{BD},\,\alpha_\mathrm{MS})$, the
number and the mass fractions of the various lens components change accordingly.

For the disc  mass function we closely follow 
\cite{kroupa02,kroupa07}, with a power law with slopes
$0.3,\,1.3,\,2.3$ in the mass ranges $((0.01-0.08),\,(0.08-0.5),\,
(0.5-1.0))~\mathrm{M}_\odot$ (the low value in the brown
dwarf region is in agreement with \citealt{allen05}), and normalisation 
$\int_{0.687}^{0.891}~\xi(\mu)~\mathrm{d}\mu=~5.9~10^{-3}~\mathrm{stars}~\mathrm{pc}^{-3}$.
To account for the remnant contributions we use the density values
reported in \cite{chabrier03}, to obtain 
$\Sigma^\mathrm{rem}=3~\mathrm{M}_\odot~\mathrm{pc}^{-2}$. 
This fixes the overall local density
for our fiducial disc model to $4.4~10^7~\mathrm{M}_\odot~\mathrm{kpc}^{-3}$.

%----------------------------------
\section{Analysis: the microlensing quantities} \label{sec:anaml}
%----------------------------------

Our main tool of investigation is the  rate of microlensing events $\Gamma$,
that carries the information of the number of events per time interval,
whereas the microlensing optical depth, $\tau$, is the instantaneous probability of a star
being magnified above a given threshold (e.g. \citealt{roulet97}).
Through the analysis of the differential rate,
given the efficiency of the experiment, one can analyse the distribution of the relevant
microlensing parameters as well as evaluate the number of expected 
microlensing events.

The microlensing rate \citep{derujula91,griest91} depends, for both sources and lenses, 
on the density and velocity distributions, on the lens mass function
and on the microlensing configuration.
Once the theoretical expression for the differential rate
is obtained  (Appendix~\ref{sec:app1}), to compare
with the results of a given experiment, 
we still need to specify the efficiency of the analysis,
usually provided as a function of the microlensing timescale
together with the value of the maximum impact parameter allowed.

Throughout the paper we will only consider the simpler microlensing event configuration,
point-mass lens and source with uniform relative motion
between lens and source, the so-called \emph{Paczy\'nski lightcurve} \citep{pacz86}. 
The effects of non-standard configuration events for the evaluation
of the microlensing quantities 
have been the object of a detailed study by \cite{glicenstein03}.
The largest changes are to be expected for binary caustic crossing
events, but these represent only a very small fraction
of the overall set so that the modifications
in the evaluated quantities should not exceed $10~\%$.

%----------------------------------
\section{Results} \label{sec:res}
%----------------------------------

%----------------------------------
\subsection{The optical depth profiles} \label{sec:tau}
%----------------------------------
The agreement among the different collaborations
(MACHO, EROS and OGLE) on the value of the observed
optical depth, and its agreement with theoretical models is, 
as already noted, a significant result of the microlensing
searches towards the Galactic bulge. 
We take advantage of this result by making the choice
to normalise the bulge central density to the
observed value of the optical depth.
As a fiducial value we take the result reported by the MACHO collaboration
towards the ``CGR'' (``Central Galactic Region'', defined in \cite{popowski05} 
as 9 out of the 94 observed fields nearest to the Galactic centre),
namely $\tau~=~2.17^{+0.47}_{-0.38}~10^{-6}$ for $(l,b)=1^\circ.50,-2^\circ.68$.
For our fiducial model, the \cite{stanek97} model E2, 
this gives us a central bulge density
of $\rho_0=9.6~10^9~\mathrm{M}_\odot~\mathrm{kpc}^{-3}$, corresponding to 
a bulge mass out to $2.5~\mathrm{kpc}$ of $1.5~10^{10}~\mathrm{M}_\odot$
(this is strictly the mass due to possible lenses). For 
the model G2 we obtain instead  $\rho_0=2.4~10^9~\mathrm{M}_\odot~\mathrm{kpc}^{-3}$
and a mass of $1.4~10^{10}~\mathrm{M}_\odot$.

\begin{figure}
\resizebox{\hsize}{!}{\includegraphics{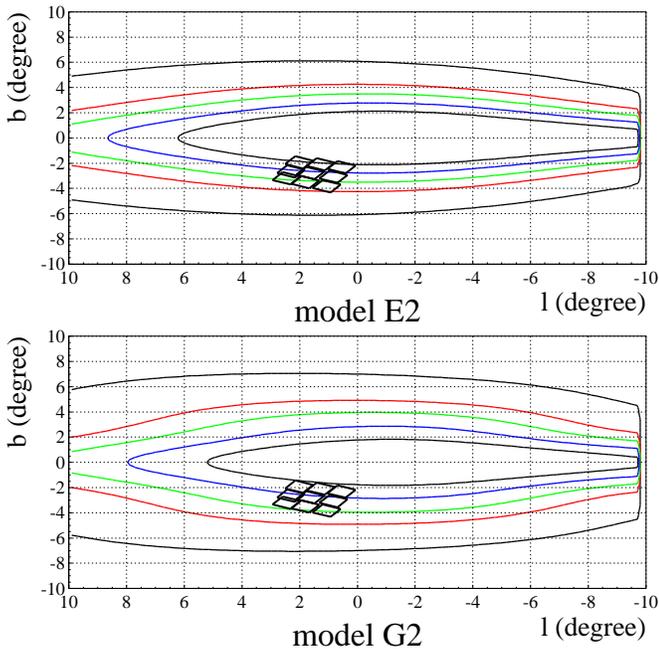}}
\caption{Optical depth profiles as a function
of the Galactic coordinates $l,b$ for the bulge models E2 (top) and G2. 
The contours of the 9 MACHO CGR fields are shown.
The optical depth is normalised to the value of the observed $\tau$
at position $l,b=1^\circ.50,-2^\circ.68$ (see text for details).
The profiles are drawn at values of $\tau=(0.3,1.0,1.5,2.17,3.0)~10^{-6}$.
Overall, the 94 MACHO fields \citep{popowski05} extend 
in the ranges $0^\circ,8^\circ$ and $-2^\circ,-10^\circ$ in Galactic 
longitude and latitude respectively.
The 66 EROS fields \citep{hamadache06} cover two regions at both positive
and negative Galactic latitude,
$l\sim(8^\circ,-6^\circ)$, $b\sim(-2^\circ,-6^\circ)$  and
$l\sim(6^\circ,-4^\circ)$, $b\sim(2^\circ,6^\circ)$. The 30 OGLE fields
analysed in \cite{sumi06} cover a smaller region near the Galactic
centre spreading only slightly beyond the MACHO CGR fields.
}
\label{fig:tau}
\end{figure}

Having normalised our model to the optical depth observed 
along a given line of sight,
next we have to test the optical depth profile
against the observed one, given that the observed events are
spread over $\sim 4^\circ$ in Galactic latitude
and $\sim 10^\circ$ in Galactic longitude.
Indeed, the optical depth profile depends strongly on
the line of sight, in particular on
the Galactic latitude (e.g. \citealt{evans02}). 
In Fig.~\ref{fig:tau} we show the optical depth
profile for the models E2 and G2. We note the larger
gradient along the Galactic latitude for the first model.

To gain further insight on their results, the EROS collaboration \citep{hamadache06}
studied the relation $\tau=\tau(b)$ giving
the empirical expression $\tau=N~\exp[-a(|b|-3^\circ)]$,
where $a,\,N$ are to be determined from the  observational data.
For the EROS data set, \cite{hamadache06} find
$N=1.62~\pm~0.23$, $a=0.43~\pm~0.16$.
As a theoretical prediction, given the EROS observational setup,
for the E2 (G2) models we find $N,a=1.77,0.52\,(1.87,0.37)$ respectively.
If we carry out the same exercise considering either
the MACHO or the OGLE observational setup we find the values
$N,a=1.60,0.56\,(1.72,0.39)$ and $N,a=1.81,0.51\,(1.94,0.34)$
respectively. Overall, the E2 and the G2 model
predictions are both consistent with the observed values.

\begin{figure}
\resizebox{\hsize}{!}{\includegraphics{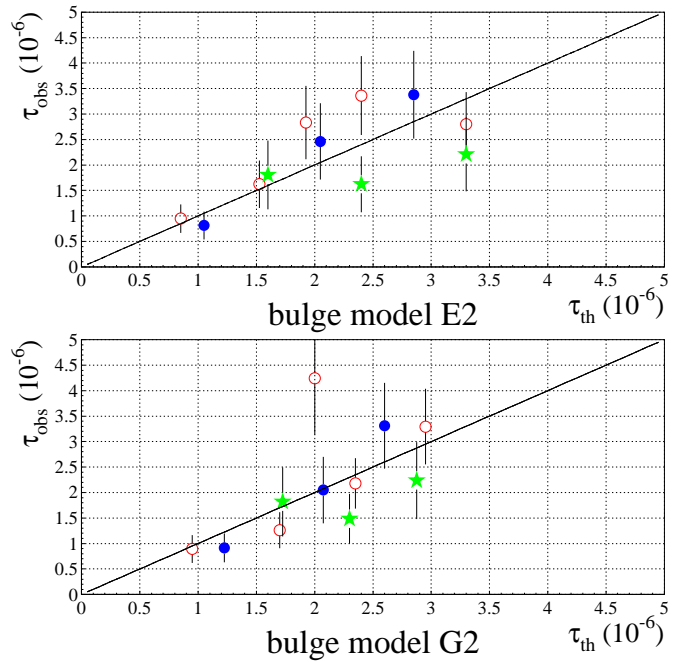}}
\caption{The observed and the expected optical depth
for the bulge models E2 (top) and G2 (see text for details).
EROS, MACHO and OGLE data are the empty, filled circles
and stars respectively. The solid line is the $y=x$ line.}
\label{fig:tau2}
\end{figure}

The previous analysis has to be carried out taking bins in the
Galactic latitude, averaging over the Galactic longitude
for the observed fields. This way, however,
one misses the information of the (albeit smoother) variation of the 
optical depth profile along the Galactic longitude.
Moreover, of course, we are comparing the expected
optical depth to the EROS observed values only.
As a different approach, we propose to take bins, instead,
in the expected optical depth, to be compared 
with the observed one as evaluated for each observational campaign. 
To perform this analysis we first need to evaluate 
the observed value of the optical depth, and therefore
the number of sources stars, in each chosen bin, whereas this number
is known \emph{per~field}. As a first order approximation
we consider the number of source stars in a given fraction
of a field to be proportional to its area.
A bin in the theoretical optical depth
delimits a region in the Galactic plane.
We choose the bin sizes so to get a (roughly) equal
number of observed events in each bin. 
For EROS (MACHO) we tried with both 5 and 10 (3 and 5) bins, resulting in
very similar results; for OGLE we use 3 bins. In Fig.~\ref{fig:tau2}
we show the observed optical depth as a function
of the expected one, for the EROS and MACHO and OGLE data sets
(with 5, 3 and 3 bins respectively)
and both models E2 and G2. 
We find very good agreement for both models
with the three data sets. 
Indeed, if we fit the relation $\tau_\mathrm{obs}=a\cdot \tau_\mathrm{th}$,
considering the 11 points of the three data sets together,
we get $a=0.9\pm0.1$, for a reduced $\chi^2=1$ for both models.

A possible way to disentangle  the different models
would come from an independent normalisation of the bulge mass.
Indeed, according to our choice, the expected optical
depth is made to coincide at the CGR MACHO location,
therefore, even if different, the two profiles
remain rather near each other along the observed fields.
The other option would be to observe events over
a larger area of the sky.
The more interesting region to be explored being
that closer to the Galactic plane.

%----------------------------------
\subsection{The Galactic Bulge IMF} \label{sec:imf2}
%----------------------------------

The microlensing rate, as discussed in Sect.~\ref{sec:anaml},
is an efficient tool for the analysis of the characteristics
of the microlensing events. Here we focus on an analysis
of the timescales provided by the current observations.
Indeed, as outlined in the Introduction, though degenerate
with other unobservable quantities (distances and relative
velocity between sources and lenses) the dependence on the lens \emph{mass}
makes the timescale a valuable source of information
on the mass function of the lens population.
Due the above-mentioned degeneracy, one needs a rather large number
of observed events to deal with them statistically.
The current observational results 
begin to provide such data set, 62 events from the MACHO collaboration
\citep{popowski05}, 120 events from the EROS collaboration  \citep{hamadache06}
and the 32 events from the OGLE collaboration \citep{sumi06}.
Note that we choose not to consider  the different data sets together,
rather, we carry out independent analyses and then compare the results.

The model, as described in the previous section, together with
the microlensing event geometry
and the experimental apparatus,
summarised in the reported detection efficiency
usually given as a function of the duration,
$\mathcal{E}=\mathcal{E}(t_\mathrm{E})$,
provides us with the expected \emph{number density}
of the microlensing events. Allowing for the Poisson
nature of the process we can write down
the likelihood \citep{gould03}, as a function of the free
parameters of our model, 
as\footnote{This is the so-called
``extended maximum likelihood'',
first proposed by Fermi (for a discussion see e.g. \citealt{barlow98}),
that is appropriate in experiments where the number of events is 
itself a random variable.}
\begin{equation} \label{eq:like}
L\left(\alpha_\mathrm{BD},~\alpha_\mathrm{MS}\right) = 
\exp(-N_\mathrm{exp})\,
\prod_{i=1}^{N_\mathrm{obs}}\left.\frac{\mathrm{d}
{\Gamma}_{i,\mathcal{E}}}{\mathrm{d}t_\mathrm{E}} 
\right|_{t_{\mathrm{E},event}}\!\!\!\!\!\!\!\!\!\!\!\!\!\!\!\!.
\end{equation}
Here $N_\mathrm{exp}$ is the overall expected
number of events, to be evaluated by integrating out
the differential rate taking into account, besides
the detection efficiency, the number of sources
and the overall duration of the experiment.
In particular it results $N_\mathrm{exp} =
N_\mathrm{exp}(\alpha_\mathrm{BD},~\alpha_\mathrm{MS})$.

As outlined in Sect.~\ref{sec:imf} we take as free parameters
the slopes of the IMF in the brown dwarfs and main sequence ranges,
$\alpha_\mathrm{BD},~\alpha_\mathrm{MS}$, that we want to estimate. 
To evaluate the likelihood, we sum the disc and the bulge contributions,
and for each the contribution of the brown dwarfs, main sequence and
remnants lens populations.

Finally, to estimate the confidence levels,
we evaluate the probability
distribution $P\left(\alpha_\mathrm{BD},~\alpha_\mathrm{MS}\right)$
by Bayesian inversion
using a flat prior on both the parameters.

It is useful, for our purposes, to take the sample of MACHO CGR
events as a ``fiducial'' sample. This provides us with a more 
homogenous, but still quite large,
set of events all located in a region small enough
to make any possible spatial dependence,
that we may not have correctly reproduced  within our model, 
almost irrelevant. Furthermore, the CGR allows a 
more straightforward comparison among the different data sets.

The 66 EROS fields \citep{hamadache06} cover a rather larger region
in the plane of the sky than the MACHO fields,
both at positive and negative Galactic latitude. 
For comparison with the MACHO CGR sample,
we select the 5 fields (5,8,607,610,611) whose location is roughly 
coincident with that of the MACHO CGR fields, and where 18 of the 120
events reported by the EROS collaboration are located.

Finally, we observe that the location of the 20 fields
used by OGLE in their analysis \citep{sumi06}
only slighly exceeds the CGR.

For the \emph{observed} distributions
for both the MACHO and the EROS data sets, there is
an increase in duration moving from the
smaller sample in the CGR to the complete data set.
In Table~\ref{tab:teobs} we report the average observed durations,
both uncorrected for the efficiency and weighted by the inverse efficiency,
the latter quantity allowing a more straightforward comparison
between the different data sets.
\begin{table}
\caption{The average observed duration, 
$\langle t_\mathrm{E}\rangle$ (days),
of the microlensing candidates
reported by the MACHO \citep{popowski05}, EROS \citep{hamadache06}
and OGLE \citep{sumi06} collaborations. In the first row,
for MACHO and EROS we report the result within the CGR 
(see text for details). For each data set, in the right column
we report the average weighted by the inverse efficiency.
}              
\label{tab:teobs} 
\centering      
\begin{tabular}{c|cc|cc|cc}
\hline\hline             
 & \multicolumn{2}{c}{MACHO} & \multicolumn{2}{c}{EROS} & \multicolumn{2}{c}{OGLE}\\    
\hline                                   
CGR &     19.5 & 15.0 & 25.9 & 22.4 & - & -\\  
all set &  28.0 & 20.0 & 32.9 & 28.3 & 32.8 & 28.1\\  
\hline
\hline
\end{tabular}
\end{table}

%----------------------------------
\subsubsection{The analysis within the CGR} \label{sec:res1}
%----------------------------------

\begin{figure}
\resizebox{\hsize}{!}{\includegraphics{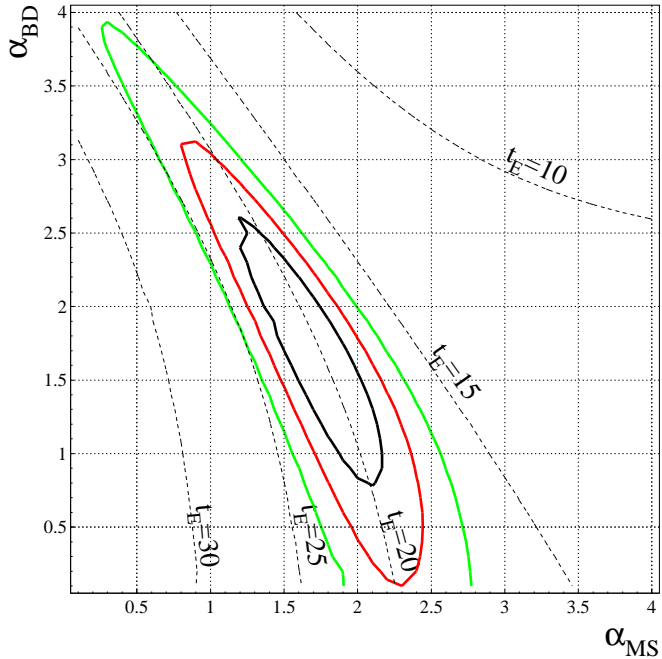}}
\caption{Probability isocontours with 34\%, 68\% and 90\% regions
in the $\alpha_\mathrm{BD},~\alpha_\mathrm{MS}$ plane.
$\alpha_\mathrm{BD},~\alpha_\mathrm{MS}$ are the slopes
of the power law IMF of the Galactic bulge lenses,
in the brown dwarf and main sequence range, respectively.
The dashed lines are the lines of equal average expected event durations,
for the values of $10,\,15,\,20,\,25$ and 30 days.
Larger values of the duration are found for smaller
values of the IMF slopes.
Here the set of 42 events reported
by the MACHO collaboration in the CGR is considered.
}
\label{fig:prob}
\end{figure}

The main result of the present paper is shown in Fig.~\ref{fig:prob}.
From the maximum likelihood analysis we show
the contours of equal probability  in the
$\alpha_\mathrm{BD},~\alpha_\mathrm{MS}$ parameter space.
Here we consider the sample of the 42 MACHO events observed within the CGR.

\begin{figure}
\resizebox{\hsize}{!}{\includegraphics{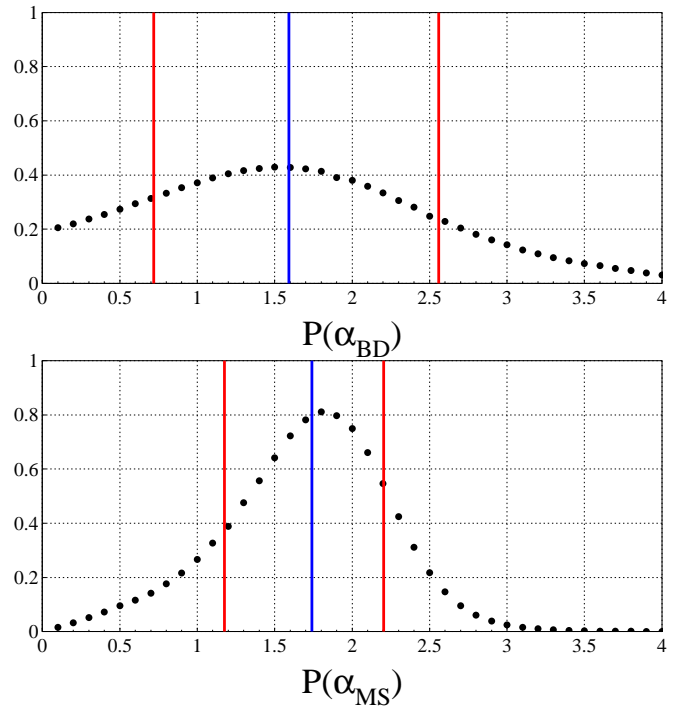}}
\caption{Probability distribution
for the Galactic bulge power law IMF slopes in the brown dwarf (top) and 
main sequence mass ranges. 
Here the set of 42 events reported
by the MACHO collaboration in the CGR is considered.
The probability lines of 16\%, 50\% and 84\% are indicated.
}
\label{fig:prob2}
\end{figure}

The  data better constrain the IMF slope in the main sequence 
range than in the brown dwarf range.
As for the IMF parameters,
at maximum probability we get the values
$\alpha_\mathrm{BD}=1.6,\,\alpha_\mathrm{MS}=1.7$.
The corresponding bulge \emph{mass} fractions are 
$\sim~( 21\%, 56\%, 17\%, 4\%, 3\%)$
for brown dwarfs, main sequence, white dwarfs, neutron stars and black
holes, respectively, for an average mass of $0.1~\mathrm{M}_\odot$.
Note the rather high brown dwarf fraction,
indeed within the $34\%$ level it does not decrease below $\sim~20\%$.
Overall the bulge contributes about $80\%$ of the events (this result
confirming the statement made about the only relative 
importance of the disc contribution, Sect.~\ref{sec:disc})
and the \emph{event} fractions due to the different lens populations are
$\sim~(29\%,57\%,11\%,2\%,1\%)$.
In Fig.~\ref{fig:prob2},
we show the one dimensional probability profile
$P\left(\alpha_\mathrm{BD}\right)$ and $P\left(\alpha_\mathrm{MS}\right)$.
As already mentioned, the $\alpha_\mathrm{MS}$ distribution 
turns out to be more peaked, with
$\alpha_\mathrm{MS}=1.7\pm0.5$ and $\alpha_\mathrm{BD}=1.6\pm1.0$.

In Fig.~\ref{fig:prob} we show the lines of equal value 
of the expected duration $t_\mathrm{E}$
superimposed on the likelihood probability contours.
As it is apparent from the plot, the lines of degeneracy in the 
parameter space $\alpha_\mathrm{MS}-\alpha_\mathrm{BD}$
that are found in the probability contours are driven by the duration
(within the innermost 34\% probability contour
the dispersion of the expected duration is only  about 5\%).
In particular, we observe that expected shorter durations
are associated with steeper mass function.
This is expected, of course, because of the relationship
between the duration and the mass of the lens.
This correlation is relevant in order to properly understand
the variations we find in the evaluated slopes of the mass function 
for either sets of data with different duration distributions or for 
different models.

\begin{figure}
\resizebox{\hsize}{!}{\includegraphics{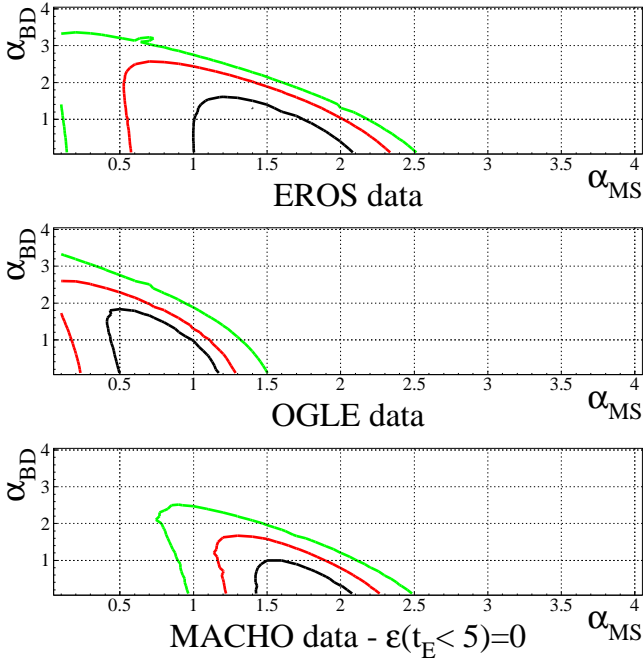}}
\caption{Probability isocontours with 34\%, 68\% and 90\% regions
in the $\alpha_\mathrm{BD},~\alpha_\mathrm{MS}$ plane.
$\alpha_\mathrm{BD},~\alpha_\mathrm{MS}$ are the slopes
of the power law IMF of the Galactic bulge lenses,
in the brown dwarf and main sequence ranges, respectively.
From top to bottom,  the results of the analysis for the EROS and OGLE data 
and the results of an analysis of the MACHO data 
where we set the efficiency below $t_\mathrm{E}<5~\mathrm{d}$ to zero
(see text for details).
For the MACHO and EROS data sets we restrict
the analysis to the subset of events observed in the inner Galactic region.
}
\label{fig:prob3}
\end{figure}

We now compare the results we obtain using the sample
of MACHO microlensing candidates \citep{popowski05} with that of the EROS
\citep{hamadache06} and of the OGLE \citep{sumi06} collaborations.
Following the previous discussion,
we first consider the samples restricted to the inner Galactic region,
namely we use 18 out of the 120 EROS microlensing candidates and the full 
set of the 32 OGLE microlensing candidates. 

As shown in Fig.~\ref{fig:prob3} (top panel), the analysis
over the EROS data set allows us to determine the maximum
for the IMF slope in the main sequence region,
roughly consistent with that found using the MACHO data set,
but does not reveal any lower limit in the brown dwarf range.
This arises because of the different 
distribution of the observed timescale.
In particular, the explanation may be traced back to the lack 
(already noted in \citealt{hamadache06}) of very 
\emph{short duration} events, say below 5 days,
within the EROS data set (both in the restricted sample of 18 events
we consider here and in the full data set).
While this difference does not significantly affect 
the results on the optical depth, in the present analysis
this turns out to be very relevant. 
The analysis performed on the OGLE data 
set provides a qualitatively similar result.
In agreement with the previous discussion, we note that the somewhat lower value for 
$\alpha_\mathrm{MS}$ is a consequence of the higher average observed timescale.

The above analysis clearly shows the extent to which
short duration events are 
essential to constrain the lower tail of the IMF.
We further address this question,
also to compare the results we obtain
with the different data sets, 
by mean of the following analysis.
We set to zero the efficiency below
a given threshold, in particular 
$\mathcal{E}(t_\mathrm{E}<5~\mathrm{d}) = 0$ and
at the same time we exclude from the analysis
those observed events with $t_\mathrm{E}<5~\mathrm{d}$,
namely, the 6 events from the MACHO sample. 

The likelihood contours we obtain for
the MACHO data set are shown in the bottom panel
of Fig.~\ref{fig:prob3}. Comparing with Fig.~\ref{fig:prob2}
we see that, as for the EROS and the OGLE data sets,
the brown dwarf slope is no longer bounded
at its lower end, while the main sequence one
peaks roughly in the same region.
Carrying out this analysis for the EROS data set
we find an almost identical result to  that
shown in the top panel of Fig.~\ref{fig:prob3},
while for OGLE we find a somewhat different
behaviour. In that case  a lower bound
for $\alpha_\mathrm{BD}$ appears,
at least for the innermost 34\% contour,
but at the same time the contours become
unbounded at the upper end. A similiar
behaviour is also observed when we move the lower limit on the lens 
mass from 0.01 to 0.04 M$_\odot$, Sect.~\ref{sec:res3},
and this can be understood, as we are excluding
the duration range where the microlensing rate 
of very large brown dwarf slopes peak.

%----------------------------------
\subsubsection{The analysis of the complete data set} \label{sec:res2}
%----------------------------------

\begin{figure}
\resizebox{\hsize}{!}{\includegraphics{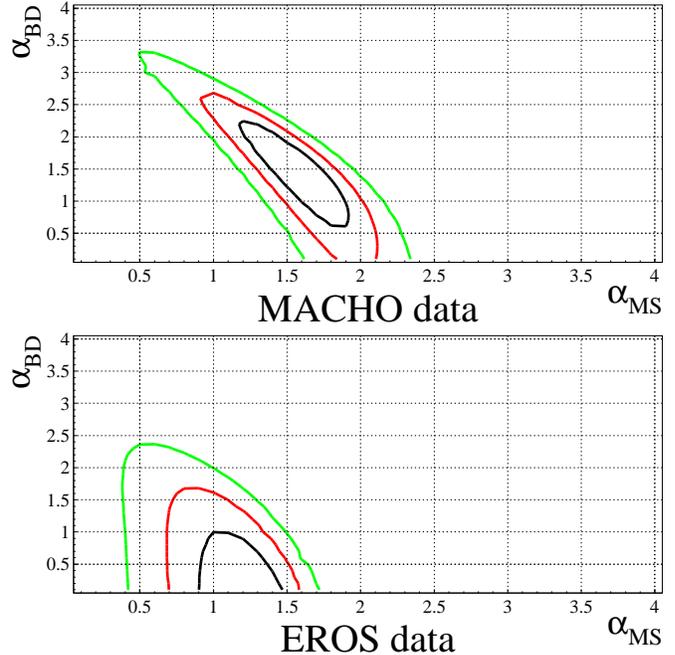}}
\caption{
Probability isocontours with 34\%, 68\% and 90\% regions
in the $\alpha_\mathrm{BD},~\alpha_\mathrm{MS}$ plane.
$\alpha_\mathrm{BD},~\alpha_\mathrm{MS}$ are the slopes
of the power law IMF of the Galactic bulge lenses,
in the brown dwarf and main sequence ranges, respectively.
The full set of events for MACHO (top) and EROS data sets are considered.
}
\label{fig:prob4}
\end{figure}

The analysis of the complete data sets 
confirm our previous conclusions.
In Fig.~\ref{fig:prob4} we show
the probability contours for both the full set of events
of MACHO and EROS (for OGLE, the results obtained
with the full data set are shown in Fig.~\ref{fig:prob3}).
For the MACHO data set
we evaluate the slope in the main sequence range to be
$\alpha_\mathrm{MS}=1.6\pm 0.4$, in agreement
with the previous result.
With respect to Fig.~\ref{fig:prob} and Fig.~\ref{fig:prob3}
we  observe, however,
the maximum likelihood contours moving  towards somewhat 
smaller values of the IMF slopes.
This is of course to be attributed to the increase
in the observed duration (Table \ref{tab:teobs}).
Comparing to Fig.~\ref{fig:prob3},
for both MACHO and EROS data sets
we observe a shrinking
in the probability contours
due to the much larger sample of events used in the present analysis.

%----------------------------------
\subsubsection{The IMF: a test against the fiducial model} \label{sec:res3}
%----------------------------------

We now focus on the possible systematic effects resulting from
a change in the characteristics of our fiducial model (Sect.~\ref{sec:models}). 
We carry out this analysis using the MACHO data set only within the CGR.
To give an indication of the goodness of the model we use
a Kolmogorov-Smirnov (KS) test comparing the expected
to the observed duration distribution (we report its significance level,
$ks$, as a disproof of the null hypothesis that the distribution
are the same, such that a low value of $ks$ indicates a poor agreement
between the expected and the observed distribution).
Such an analysis makes sense because  the variations in the expected 
timescale do not exceed $\sim~5\%$ across the CGR.
This allows us to carry out the KS test by evaluating 
an \emph{average} rate, summing 
the rate observed towards the different fields with a weight given
by the number of source stars that we compare to the observed duration distribution.
As test models we consider the following (Table \ref{tab:res}):
model 2 : we change the bulge dispersion velocity to the \cite{rattenbury07b} values
(Sect.~\ref{sec:vel});   
model 3-4 : we change respectively downward and upward the bulk rotation velocity of the bulge 
(Sect.~\ref{sec:vel});
model 5 : we change the parameters of the disc density profile according 
to \cite{hangould03} (Sect.~\ref{sec:disc});
model 6 : we change the disc density profile according 
to \cite{freudenreich98} (Sect.~\ref{sec:disc});
model 7 : we change the bulge scale lengths according to \cite{rattenbury07a} 
(Sect.~\ref{sec:bulge});
model 8 : we change the bulge IMF slope to $\alpha_\mathrm{rem}=2.7$ (Sect.~\ref{sec:imf});
model 9 : we change the bulge mass function introducing a second slope in the main sequence
range (Sect.~\ref{sec:imf});
10) we change the mass lower limit in the brown dwarf range to $0.04~\mathrm{M}_{\odot}$
(Sect.~\ref{sec:imf}).
In Table~\ref{tab:res} we report the results: 
for each model we give the evaluated $\alpha_\mathrm{MS}$
parameter out of the $P\left(\alpha_\mathrm{MS}\right)$ distribution
(with the 16\%, 50\% and 84\% bound) and the KS significance level.

\begin{table}
\caption{Results of the maximum likelihood analysis 
on the MACHO data set \citep{popowski05}
for our fiducial model and 
for the different models discussed (Sect.~\ref{sec:models}).
$\alpha_{MS}$ is the slope of the power law IMF
in the main sequence range. 
$ks$ is the Kolmogorov-Smirnov significance level for the null hypothesis that
the expected and observed distributions are the same.
}              
\label{tab:res} 
\centering      
\begin{tabular}{c|ccc|cc}
\hline\hline             
model & \multicolumn{3}{c}{$\alpha_{\mathrm{MS}}$} & $ks$ & model change\\    
 & 16\% & 50\% & 84\% &&\\
\hline                                   
    1 & 1.18 & 1.74 & 2.20 & 0.39 & fiducial model\\  
\hline                                  
    2 & 0.86 & 1.40 & 1.80 & 0.29 & bulge velocity  \\
    3 & 1.16 & 1.73 & 2.19 & 0.39 & bulge velocity  \\ 
    4 & 1.17 & 1.74 & 2.20 & 0.39 & bulge velocity  \\
    5 & 1.32 & 2.10 & 2.90 & 0.41 & disc model \\
    6 & 1.13 & 1.67 & 2.10 & 0.40 & disc model \\   
    7 & 0.97 & 1.51 & 1.92 & 0.33 & bulge model \\  
    8 & 0.95 & 1.56 & 2.06 & 0.40 & mass function\\
    9 & 0.90 & 1.62 & 2.21 & 0.40 & mass function\\
   10 & 1.86 & 2.19 & 2.60 & 0.26 & mass function\\ 		
\hline
\hline
\end{tabular}
\end{table}

For the different models the likelihood maximum moves on the 
$\alpha_\mathrm{BD}$-$\alpha_\mathrm{MS}$ plane so as to always peak  around 
the same expected timescale, with the
resulting mass function slopes changing accordingly.
The largest variation downward, $\alpha_\mathrm{MS}\sim~1.4$,
is found for model 2 as an effect
of the increased bulge velocity dispersions.
Note the large value we obtain for model 5, we find $\alpha_\mathrm{MS}\sim~2.1$.
Here two different effects push in the same direction towards
a steeper mass function, namely a smaller bulge contribution and a decrease in the
bulge dispersion velocity.
The qualitative shape of the likelihood contours
does not change for any of the models except the last.
Here, as an effect of the increase of the minimum mass value
in the brown dwarf range, from 0.01 to 0.04 M$_\odot$,
the probability distribution for $\alpha_\mathrm{BD}$
becomes unbounded at its upper end.
Correspondingly, we also find a steeper mass function
and lower KS significance level.
Overall, the variations we find for $\alpha_\mathrm{MS}$
for the different models we have tested
do not exceed the statistic uncertainty we have in our fiducial configuration.
This is in agreement with the KS analysis, according to which
we obtain acceptable results for all the models we consider.

\begin{figure}
\resizebox{\hsize}{!}{\includegraphics{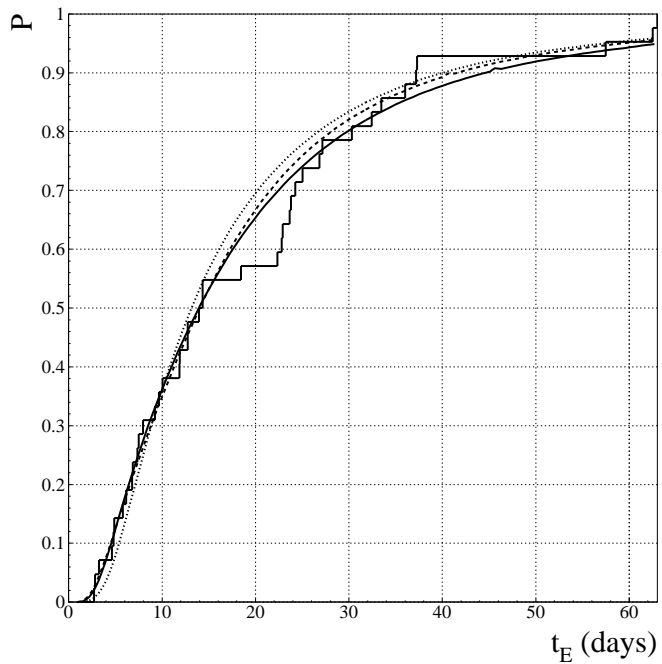}}
\caption{Cumulative duration distributions (P)
for the observed events, together with the
the theoretical prediction for three different models. 
Solid, dashed and dotted lines are for the fiducial model 
and models 2 and 10, respectively
(see text for details). Here the set of 42 events reported
by the MACHO collaboration in the CGR is considered.
}
\label{fig:ks}
\end{figure}

In Fig.~\ref{fig:ks} we show the cumulative distribution for the
sample of the 42 MACHO CGR events together with the theoretical
cumulative distributions for the fiducial model 
and models 2 and 10, for which we obtain the smallest and the largest
values for the main sequence slope ($\alpha_\mathrm{MS}=1.4,\,2.2$,
respectively) and the worst agreement 
according to the KS test. Besides the lack 
of observed events at $t_\mathrm{E}~\sim~20~\mathrm{d}$,
we note in particular 
the very good agreement with short duration events
for both the fiducial model and model 2 and the better agreement
with long duration events for model 10.

%----------------------------------
\subsection{The expected number of microlensing events} \label{sec:nexp}
%----------------------------------

Besides the study of the duration distribution,
the analysis of the microlensing rate allows one to evaluate
the number of expected events.
Through the analysis we have
normalised the bulge central density, once having
fixed that of the disc, by using the observed value
of the optical depth. Because of the relationship
between the optical depth and the microlensing rate,
through the event duration, we may therefore expect to find
a good agreement between the observed and the expected number 
of microlensing events. 
Indeed, even if the number of expected events 
varies by almost a factor of 3 across the  
$\alpha_\mathrm{BD}-\alpha_\mathrm{MS}$ parameter space we explore,
we  find a fair agreement.
For the MACHO data set our prediction is compatible
within $1~\sigma$ to the observed value; we find 38
and 54 events compared
with 42 and 62 events, in the CGR and the complete data set, respectively.
For the EROS and OGLE data sets we arrive at an even better agreement,
with an expected number of 118 (31) compared with 120 (32).
These figures do not vary significantly
(at most by $\sim 2$ events) within the innermost 34\% probability contour.

%----------------------------------
\subsection{The blending issue} \label{sec:blending}
%----------------------------------

In very crowded fields, such as those observed towards the Galactic centre, 
the observed objects can be the blend 
of several stars. This \emph{blending} effect  
is a major source of concern
for the interpretation of microlensing searches.
This is the reason, we recall,
that led, in evaluating the optical depth, to the choice of considering 
only bright sources for which
one expects the blending effects to be alleviated.
The multiple effects of blending are 
supposed to roughly balance each other when evaluating
the optical depth (see e.g. \citealt{hamadache06}).
On the other hand, as blending is expected to cause
an \emph{underestimation} of the evaluated event duration,
we may question its relevance with respect
to our results.

The extent to which blending contaminates the sample of red clump giants
is a subject of debate \citep{popowski05,sumi06,hamadache06,smith07}. 
For the present analysis
we remark that both MACHO and EROS evaluate the optical depth
without including the effect of blending, while OGLE,
who find this effect to be relevant within their data set,
use blended fits. Throughout our analysis we have used the reported
values  of the duration according to this choice.

\cite{popowski05}, for the MACHO collaboration, identify
an ``extremely conservative'' subset of events
for which they evaluate the blend fraction to be very close to 1.
We carry out our analysis on this subsample,
composed of 22 events within the CGR.
As in \cite{popowski05} we introduce an overall
normalisation factor for the microlensing rate
equal to the ratio of the number of events in this restricted
sample  to that of the complete sample.
This is coherent with the purpose of the analysis,
where one wants to test whether blending substantially
affects the event parameters,
and the derived quantities such as the optical depth
and the microlensing rate, while assuming
that it does not change the number of detected events.
As a result we find somewhat broader contours,
because of the smaller number of events, 
with the brown dwarf slope unbounded at its lower tail,
but otherwise fully compatible with our previous results.
This is in agreement with our previous discussion.
Indeed, the average observed
duration for this sample turns out to be
similar to the full CGR sample $\langle t_\mathrm{E}\rangle=22.3\,\mathrm{d}$
but 5 out of the 6 very short duration events are excluded.

On the other hand \cite{sumi06}, for the OGLE collaboration, worked the other 
way round. They repeated  their analysis assuming
no blending, finding a new sample of 48 microlensing candidates,
with an average duration roughly 20\% shorter than in the
32 events sample. 
Within this new sample there are 3 candidates with 
$t_\mathrm{E}<2~\mathrm{d}$, and this is of course relevant
in the view of our previous discussion.
However, these candidates are strongly affected by blending.
We prefer, therefore, not to include them in our analysis.
Our likelihood analysis carried out on this 45-event subsample
turns is compatible with the previous one.

Finally, we recall that EROS \citep{hamadache06} conclude that blending
does not affect significantly their results, and comment
the apparent discrepancy with the result obtained by OGLE
on this issue on the basis of their different choice
for the threshold value of the amplification (in particular,
OGLE consider also less amplified events with respect to both
EROS and MACHO analyses).

In conclusion, given the available data sets, blending,
though relevant, 
does not appear to  significantly affect our results.

%----------------------------------
\subsection{Long duration events} \label{sec:long}
%----------------------------------

\begin{figure}
\resizebox{\hsize}{!}{\includegraphics{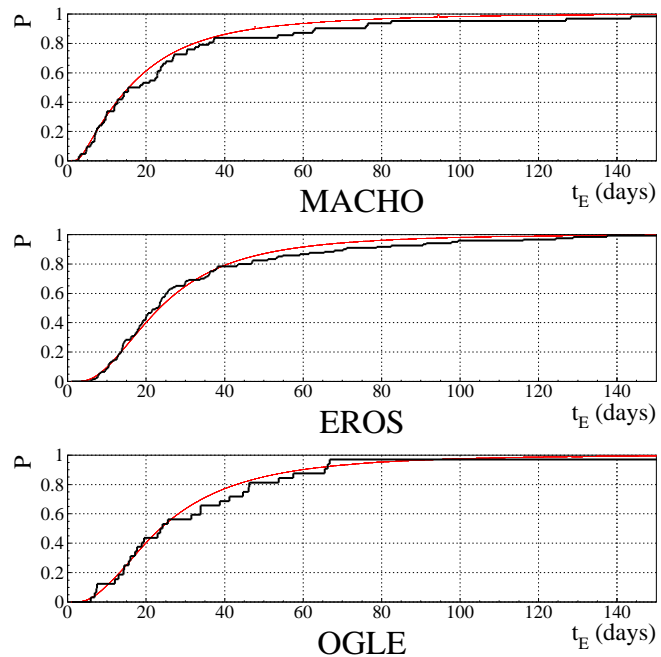}}
\caption{Cumulative duration distributions (P)
for (top to bottom) the MACHO, EROS and OGLE
sample of events. Superimposed, the predicted
distribution for the fiducial model.
}
\label{fig:cumul}
\end{figure}

In Fig.~\ref{fig:cumul} we show, for the three complete sets of events
we consider (MACHO, EROS and OGLE), the cumulative duration distribution
and the expected cumulative distribution
for the fiducial model (averaged as for the KS analysis in Sect.~\ref{sec:res3}),
evaluated at the IMF slopes that maximise the likelihood.
As for the smaller CGR sample in Fig.~\ref{fig:ks},
we note the rather good agreement especially for short duration events.
For both EROS and MACHO data sets we also observe a systematic
excess of long duration events. This turns out to be, however,
only marginally significant. For instance, the models
predict 10\% of events with $t_\mathrm{E}>51~\mathrm{d}$,
compared with $\sim~15\%$ of the observed events.

%----------------------------------
\section{Conclusions} \label{sec:end}
%----------------------------------

We have considered
the sample of microlensing events observed
towards the Galactic bulge with red clump giant sources 
reported by the  the MACHO \citep{popowski05}, OGLE  \citep{sumi06} and 
EROS \citep{hamadache06} collaborations
to place constraints 
on the bulge mass function. 
In particular, through a likelihood analysis,
we have studied the slopes, $\alpha_\mathrm{BD},\alpha_\mathrm{MS}$,
of a power law mass function in the brown dwarf $(0.01-0.08)~\mathrm{M}_\odot$
and main sequence $(0.08-1.)~\mathrm{M}_\odot$ mass ranges.

For our fiducial model, comparing to the CGR sample
of 42 MACHO events, we obtain $\alpha_\mathrm{MS}=1.7\pm 0.5$.
This result compares well to that obtained in the $(0.15-1)~\mathrm{M}_\odot$
range by \cite{zoccali00}, $\alpha\sim 1.3$. 
The slope in the brown dwarf range turns out to be less well
constrained, $\alpha_\mathrm{BD}=1.6\pm 1$. 
Overall our maximum likelihood results indicate a rather significant
contribution of low mass lenses, with $\sim~30\%$ of the events to be
attributed to brown dwarfs. The last result is in agreement
with previous analyses \citep{hangould96,grenacher99}, while
our value for the main sequence slope, somewhat smaller,
may be explained because we are using a more suitable
sample of red-clump-source events.

The analyses of the EROS and OGLE data sets give us
somewhat different results. We derive a smaller value
for the slope in the main sequence range, although
compatible with the MACHO data set result, but we
obtain only an upper limit for the slope of the
brown dwarf population. This behaviour finds its explanation
in the different observed timescales. In particular,
very short timescale events
($t_\mathrm{E}<5~\mathrm{d}$), only observed in the MACHO data set, 
are essential to constrain the brown dwarf mass function.
The lack of short timescale events has been noticed
in the EROS analysis \citep{hamadache06}. In all of the experiments, 
the detection efficiency of short durations events is
extremely low, 
rendering the analysis in the brown dwarf regime difficult and, therefore, 
making the result less robust.

More reliable constraints on the mass function
may come from a better understanding of the bulge model,
but especially, as already stressed , by improving the statistics
of observed short duration events.

Furthermore, we have carried out an analysis on the optical depth.
The agreement with the expected values is 
recognized \citep{hangould03}. Here we have considered
the profile of the expected optical depth as compared
to the observed one, finding a good agreement
for both the models we have considered, the model E2 of \cite{stanek97} and
the model G2 of \cite{dwek95}. To further constrain the bulge profile
it would be useful to extend microlensing searches to cover a larger area in the sky plane.

\begin{acknowledgements}
We are grateful to the referee, A. Gould, 
for useful comments and suggestions.
We thank K. Griest, C. Hamadache, A. Milsztajn and T. Sumi 
of the MACHO, EROS and OGLE collaborations 
for providing the efficiency tables and other data in electronic format.
SCN thanks Jean Kaplan for valuable discussions.
SCN, LM and GS acknowledge support for this work by MIUR through 
PRIN 2006 Prot. 2006023491\_003
and by research funds of the Salerno University.
FDL work was performed under the auspices of the EU, which has provided
financial support to the ``Dottorato di Ricerca Internazionale in
Fisica della Gravitazione ed Astrofisica'' of Salerno University,
through ``Fondo Sociale Europeo, Misura III.4''.
FDL acknowledges the Forschungskredit of the University 
of Z\"urich for financial support.
\end{acknowledgements}

\begin{appendix}

\section{Evaluation of the microlensing rate} \label{sec:app1}

In this appendix we detail the evaluation of the differential
microlensing rate (Sect.~\ref{sec:anaml}).
This quantity is directly related to
the number of expected microlensing events evaluated as
\begin{equation} \label{eq:dnev}
\mathrm{d}N_{ev}=N_{obs}T_{obs} \mathrm{d}\Gamma
\end{equation}
where $\mathrm{d}\Gamma$ is the \emph{differential rate} at which a single star 
is microlensed, $N_{obs}$ is the number of monitored sources and $T_{obs}$ 
is the whole observation time. 

The microlensing rate expresses the number of lenses that pass through 
the microlensing tube $\mathrm{d}^3x$ in the time interval $\mathrm{d}t$,
for a given number density distribution $n_l(\vec{x})$ 
and velocity distribution $f(\vec{\tilde v}_l)$ of the lenses. 
It reads \citep{derujula91,griest91}
\begin{equation} \label{eq:rate0}
\mathrm{d}\Gamma=\frac{n_l(\vec{x})\mathrm{d}^3x} {\mathrm{d}t} 
\times  f(\vec{\tilde v}_l) \mathrm{d}^3 {\tilde v}_l
\times \frac{\rho_s D_s^{2-\gamma} \mathrm{d}D_s}{I_s}\times
f( \vec{\tilde v}_s) \mathrm{d}^3 \tilde v_s.
\end{equation}
The two last terms account respectively for the spatial and velocity
distribution of the sources, with $I_s = \int \rho_s D_s^{2-\gamma}\ dD_s$.
We take into account that the volume element varies with distance as $D_s^2 dD_s$, $D_s$
being the distance between observer and source, and that the fraction
of monitored stars having a luminosity higher than a minimum detectable luminosity,
$L_{*}$, scales as $L_{*}^{-\gamma/2}\sim D_s^{-\gamma}$ \citep{kiraga94}.
Throughout the analysis of this paper we use $\gamma=0$,
characteristic for bright stars that can be considered,
at least approximately, as standard candles.

The volume element of the microlensing tube is 
$\mathrm{d}^3 x=(\vec{v}_{r,\bot} \cdot \hat{\vec{n}}) \mathrm{d}t \mathrm{d}S$.
$\mathrm{d}S=\mathrm{d}l\mathrm{d}D_l$  is the portion of the tube external surface,
$D_l$ is the distance between observer and lens,
and $\mathrm{d}l=u_t R_\mathrm{E} \mathrm{d}\alpha$, 
where $R_\mathrm{E}$ is the lens Einstein radius,
$u_t$ is the maximum impact parameter, 
$\vec{v}_{r,\bot}$ is the component of lens velocity in the plane
orthogonal to the line of sight (hereafter \emph{los}),
and $\hat{\vec{n}}$ is the unit vector normal to the tube inner surface 
at the point where  the microlensing tube is crossed by the lens.
In the following  $\theta$ is the angle 
between $\vec{v}_r$ and $\hat n$, with $\theta\in(-\frac{\pi}{2},\frac{\pi}{2})$
(one considers only lenses that enter  the tube).

The velocity of the lenses entering the tube reads
\begin{equation} \label{eq:vtube}
\vec{\tilde v}_l = \vec{v}_r+\vec{v}_t,
\end{equation}
where $\vec{v}_t$ is the tube velocity. 
On the lens plane, we have $\vec{v}_t=(1-x) \vec{v}_{obs}+x \vec{\tilde v}_s$, where
$x\equiv\frac{D_l}{D_s}$, $\vec{v}_{obs}$ is the observer's (solar)
velocity  and $\vec{\tilde v}_s$ is the source velocity.
We decompose both lens and source velocities into a random
plus a bulk component $\vec{\tilde v}=\vec{v}+\vec{v}_{drift}$. 
In conclusion
\begin{equation} \label{eq:vl}
\vec{v}_{l}=\vec{v}_{r}+x\vec{v}_{s}+\vec{A},
\end{equation}
where we have defined the vector $\vec{A}$ so as to include all the bulk motion,
$\vec{A}\equiv (\vec{v}_{obs}-\vec{v}_{drift,l})+x (\vec{v}_{drift,s}-\vec{v}_{obs})$.
For the random component we use an anisotropic Gaussian distribution (Sect.~\ref{sec:vel}).

Looking at Eq.~\ref{eq:rate0} we see that we start from the joint three-dimensional
velocity and source distributions. However, only the distribution
of the relative velocity on the lens plane is relevant to the 
microlensing rate, since it determines the lensing time scales $t_\mathrm{E}$
\emph{via} the relation $t_\mathrm{E}=R_\mathrm{E}/v_r$. As we show below,
it is possible to analytically evaluate this distribution 
\citep{arno06}.
Indeed, besides the velocity components along the \emph{los}, both
the remaining components of the source velocity can be analytically integrated.
A final integration on the remaining component of the lens velocity
is not possible, however, as a consequence of the 
assumed anisotropy of the velocity distribution.

The rationale of the evaluation is as follows. A first integration
along the \emph{los} for both sources and lenses leaves us
with two Gaussian non diagonal distributions, that we project
on the lens plane and diagonalise. This defines two frames  on the lens plane
whose axes, in general, will be misaligned.
We then fix one of the two frames as a ``reference'',
in particular that relative to the lens velocity distribution,
we evaluate the relative velocity distribution by making use
of Eq.~\ref{eq:vl}  and  integrate out
the source velocity distribution components.

After integration along the line of sight the two-dimensional
velocity distribution on the lens plane orthogonal to the \emph{los} is
\begin{equation} \label{eq:pv2d}
f(\vec{v}_{(i)p})  \mathrm{d}^2 v_{(i)p}=
\frac{\mathrm{e}^{-\frac{v^2_{(i)p,1}}{2\sigma^2_{(i)p,1}}}
\mathrm{e}^{-\frac{v^2_{(i)p,2}}{2\sigma^2_{(i)p,2}}}}
{(2\pi)^\frac{3}{2}\sigma_{(i)p,1}\sigma_{(i)p,2}}\,\mathrm{d}^2 v_{(i)p}\,,
\end{equation}
where the suffix $(i)$ indicate either lenses or sources. For the last component,
because of the projection, $\vec{v}_{sp}=x~\vec{v}_s$
and $\sigma_{sp\{1,2\}} =  x~\sigma_{s,\{1,2\}}$. 

\begin{figure}
\begin{center}
\includegraphics[width=7cm, height=5cm]{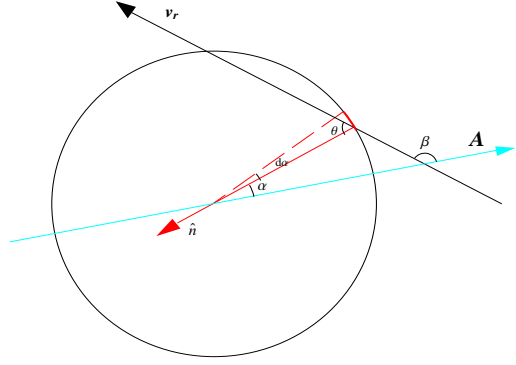}
\caption{
The microlensing tube cross section and the angles involved.}
\label{fig:angoli}
\end{center}
\end{figure}

Let the principal axes of the intersection ellipse of the lens (source)
proper velocity ellipsoid with the lens plane be $\{x_{l,1},x_{l,2}\}$ 
($\{x_{s,1},x_{s,2}\}$), hereafter we refer to the former frame as $OL$,
$\omega$ the angle between $x_{l,1}$ and $x_{s,1}$ 
and $\{v_{spl,1},v_{spl,2}\}$ the source velocity components in $OL$.
$\omega$, as well as the values for the projected dispersion velocities,
are fixed by the geometry, once the \emph{los} has been chosen
($\omega$ varies up to $\sim~8^\circ$, 
increasing with the Galactic latitude).
The distribution for the relative velocity is then evaluated as
\begin{eqnarray}
\label{eq:pvel}
&&P(\vec{v}_r) \mathrm{d}^2 {v}_r = \int  P(\vec{v}_r,\vec{v}_{spl}) \mathrm{d}^2 v_{spl}
\mathrm{d}^2 v_r = 
\nonumber \\
&& \int\!\!f(\vec{v}_{spl})f(\vec{v}_{lp})\delta(\vec{v}_r\!-\!(\vec{v}_{lp}-\vec{v}_{spl}+\vec{A})) 
\mathrm{d}^2 v_{lp} \mathrm{d}^2 v_{spl} \mathrm{d}^2 {v}_r\!\!= 
\nonumber \\
&&  \frac{1}{\pi \Sigma_N} 
\mathrm{e}^{-(\vec{v}_r+\vec{A})\cdot \Sigma \cdot (\vec{v}_r + \vec{A})} 
\mathrm{d}^2 {v}_r\,,
\end{eqnarray}
where
\begin{eqnarray}
\Sigma & = & \frac{1}{\Sigma_N^2}
\left[
\begin{array}{cc}
\Sigma_a+\Sigma_b & 0 \vspace{5 pt} \\
0 & \Sigma_a-\Sigma_b\vspace{5 pt} \\
\end{array}
\right]\,,\nonumber\\
\Sigma_a & = &  \sigma_{l,1}^2+\sigma_{l,2}^2+x^2 (\sigma_{s,1}^2 +\sigma_{s,2}^2), \nonumber \\
\Sigma_b & = & (4 x^2 \sin^2\omega(\sigma_{l,2}^2-\sigma_{l,1}^2) (\sigma_{s,1}^2-\sigma_{s,2}^2)+ \nonumber \\
&&(\sigma_{l,1}^2-\sigma_{l,2}^2+x^2(\sigma_{s,1}^2-\sigma_{s,2}^2))^2)^{1/2},\nonumber \\
\Sigma_N  & = & ((\sigma_{l,1}^2+x^2 \sigma_{s,1}^2)(\sigma_{l,2}^2+x^2\sigma_{s,2}^2) + \nonumber \\
&&x^2 (\sigma_{l,1}^2-\sigma_{l,2}^2)(\sigma_{s,1}^2-\sigma_{s,2}^2) \sin^2\omega)^{1/2}\,.
\end{eqnarray}
Here both $\vec{v}_r$ and $\vec{A}$ must be evaluated in $OL$. The angle between
$\vec{A}$ and $x_{l,1}$ is fixed by the geometry once we assign the \emph{los}.
Moreover, as Fig.~\ref{fig:angoli} shows, $\beta = \pi+\alpha-\theta$ is
the angle between $\vec{v}_r$ and $\vec{A}$.

Finally, after moving to polar coordinates on the lens plane, ($v_r,\theta$),
we reach the following expression for the microlensing rate
\begin{eqnarray} \label{eq:rate}
\mathrm{d}\Gamma & = &2\, f(\mu) \rho_l\, 
\frac{\rho_s D_s^{2} \mathrm{d} D_s}{\mathrm{I}_s}\,
R_\mathrm{E}(\mu,D_l,D_s) u_t \times\nonumber\\
&&P(v_r,\alpha)\,v_r^2 \mathrm{d}v_r\,
\mathrm{d}\mu\,\mathrm{d}D_l\,\mathrm{d}D_s\,\mathrm{d}\alpha\,,
\end{eqnarray}
where we have exploited the  periodicity of the trigonometric functions involved 
in $P(\vec{v}_r)$ to analytically integrate over $\theta$ (this provides the factor '2').
Because of the assumed anisotropic velocity dispersions, the
analytical integration over $\alpha$ is not possible
and the dependence on this variable survives in the relative velocity distribution.
The expression for the differential rate $\mathrm{d}\Gamma/\mathrm{d}t_\mathrm{E}$
is easily obtained from Eq.~\ref{eq:rate} using the relation $v_r={R_E}/{T_E}$.

In Eq.~\ref{eq:rate} we have introduced the dependence of the lens number
distribution on the mass of the lens, $\mu$, with the usual 
``factorisation hypothesis'' stating that the lens mass distribution is independent 
of the lens spatial distribution \citep{derujula91}. $\rho_l$ 
is the lens spatial distribution, $f(\mu)$ the lens mass function 
that we normalise as follows
\begin{equation} \label{eq:normimf}
\int^{\mu_{max}}_{\mu_{min}} f(\mu) \mu \mathrm{d}\mu=\frac{\rho_{0,l}}{M_{\odot}}\,,
\end{equation}
where $\rho_{0,l}$ is the central density.

\end{appendix}

\bibliographystyle{aa}
\bibliography{biblio}

\begin{thebibliography}{65}
\expandafter\ifx\csname natexlab\endcsname\relax\def\natexlab#1{#1}\fi

\bibitem[{{Alcock} {et~al.}(2000){Alcock}, {Allsman}, {Alves}, {Axelrod},
  {Becker}, {Bennett}, {Cook}, {Dalal}, {Drake}, {Freeman}, {Geha}, {Griest},
  {Lehner}, {Marshall}, {Minniti}, {Nelson}, {Peterson}, {Popowski}, {Pratt},
  {Quinn}, {Stubbs}, {Sutherland}, {Tomaney}, {Vandehei}, \& {Welch}}]{macho00}
{Alcock}, C., {Allsman}, R.~A., {Alves}, D.~R., {et~al.} 2000, \apj, 542, 281

\bibitem[{{Allen} {et~al.}(2005){Allen}, {Koerner}, {Reid}, \&
  {Trilling}}]{allen05}
{Allen}, P.~R., {Koerner}, D.~W., {Reid}, I.~N., \& {Trilling}, D.~E. 2005,
  \apj, 625, 385

\bibitem[{{Barlow}(1989)}]{barlow98}
{Barlow}, R. 1989, {Statistics. A guide to the use of statistical methods in
  the physical sciences} (The Manchester Physics Series, New York: Wiley, 1989)

\bibitem[{{Benjamin} {et~al.}(2005){Benjamin}, {Churchwell}, {Babler},
  {Indebetouw}, {Meade}, {Whitney}, {Watson}, {Wolfire}, {Wolff}, {Ignace},
  {Bania}, {Bracker}, {Clemens}, {Chomiuk}, {Cohen}, {Dickey}, {Jackson},
  {Kobulnicky}, {Mercer}, {Mathis}, {Stolovy}, \& {Uzpen}}]{glimpse05}
{Benjamin}, R.~A., {Churchwell}, E., {Babler}, B.~L., {et~al.} 2005, \apjl,
  630, L149

\bibitem[{{Binney} \& {Merrifield}(1998)}]{BM98}
{Binney}, J. \& {Merrifield}, M. 1998, {Galactic astronomy} (Galactic astronomy
  / James Binney and Michael Merrifield.~ Princeton, NJ : Princeton University
  Press, 1998.~ (Princeton series in astrophysics) QB857 .B522 1998)

\bibitem[{{Bissantz} {et~al.}(2004){Bissantz}, {Debattista}, \&
  {Gerhard}}]{bissantz04}
{Bissantz}, N., {Debattista}, V.~P., \& {Gerhard}, O. 2004, \apjl, 601, L155

\bibitem[{{Bissantz} {et~al.}(1997){Bissantz}, {Englmaier}, {Binney}, \&
  {Gerhard}}]{bissantz97}
{Bissantz}, N., {Englmaier}, P., {Binney}, J., \& {Gerhard}, O. 1997, \mnras,
  289, 651

\bibitem[{{Bissantz} {et~al.}(2003){Bissantz}, {Englmaier}, \&
  {Gerhard}}]{bissantz03}
{Bissantz}, N., {Englmaier}, P., \& {Gerhard}, O. 2003, \mnras, 340, 949

\bibitem[{{Bissantz} \& {Gerhard}(2002)}]{bissantz02}
{Bissantz}, N. \& {Gerhard}, O. 2002, \mnras, 330, 591

\bibitem[{{Blum}(1995)}]{blum95}
{Blum}, R.~D. 1995, \apjl, 444, L89

\bibitem[{{Cabrera-Lavers} {et~al.}(2007){Cabrera-Lavers}, {Hammersley},
  {Gonz{\'a}lez-Fern{\'a}ndez}, {L{\'o}pez-Corredoira}, {Garz{\'o}n}, \&
  {Mahoney}}]{cabrera07}
{Cabrera-Lavers}, A., {Hammersley}, P.~L., {Gonz{\'a}lez-Fern{\'a}ndez}, C.,
  {et~al.} 2007, \aap, 465, 825

\bibitem[{{Calchi Novati} {et~al.}(2006){Calchi Novati}, {De Luca}, {Jetzer},
  \& {Scarpetta}}]{novati06}
{Calchi Novati}, S., {De Luca}, F., {Jetzer}, P., \& {Scarpetta}, G. 2006,
  \aap, 459, 407

\bibitem[{{Calchi Novati} {et~al.}(2005){Calchi Novati}, {Paulin-Henriksson},
  {An}, {Baillon}, {Belokurov}, {Carr}, {Cr{\'e}z{\'e}}, {Evans},
  {Giraud-H{\'e}raud}, {Gould}, {Hewett}, {Jetzer}, {Kaplan}, {Kerins},
  {Smartt}, {Stalin}, {Tsapras}, \& {Weston}}]{novati05}
{Calchi Novati}, S., {Paulin-Henriksson}, S., {An}, J., {et~al.} 2005, \aap,
  443, 911

\bibitem[{{Chabrier}(2003)}]{chabrier03}
{Chabrier}, G. 2003, \pasp, 115, 763

\bibitem[{{de Jong} {et~al.}(2006){de Jong}, {Widrow}, {Cseresnjes}, {Kuijken},
  {Crotts}, {Bergier}, {Baltz}, {Gyuk}, {Sackett}, {Uglesich}, \&
  {Sutherland}}]{mega06}
{de Jong}, J.~T.~A., {Widrow}, L.~M., {Cseresnjes}, P., {et~al.} 2006, \aap,
  446, 855

\bibitem[{{De Rujula} {et~al.}(1991){De Rujula}, {Jetzer}, \&
  {Masso}}]{derujula91}
{De Rujula}, A., {Jetzer}, P., \& {Masso}, E. 1991, \mnras, 250, 348

\bibitem[{{Dehnen}(2000)}]{dehnen00}
{Dehnen}, W. 2000, \aj, 119, 800

\bibitem[{{Dehnen} \& {Binney}(1998)}]{dehnen98}
{Dehnen}, W. \& {Binney}, J. 1998, \mnras, 294, 429

\bibitem[{{Dwek} {et~al.}(1995){Dwek}, {Arendt}, {Hauser}, {Kelsall}, {Lisse},
  {Moseley}, {Silverberg}, {Sodroski}, \& {Weiland}}]{dwek95}
{Dwek}, E., {Arendt}, R.~G., {Hauser}, M.~G., {et~al.} 1995, \apj, 445, 716

\bibitem[{{Evans} \& {Belokurov}(2002)}]{evans02}
{Evans}, N.~W. \& {Belokurov}, V. 2002, \apjl, 567, L119

\bibitem[{{Freudenreich}(1998)}]{freudenreich98}
{Freudenreich}, H.~T. 1998, \apj, 492, 495

\bibitem[{{Glicenstein}(2003)}]{glicenstein03}
{Glicenstein}, J.-F. 2003, \apj, 584, 278

\bibitem[{{Gould}(2000)}]{gould00}
{Gould}, A. 2000, \apj, 535, 928

\bibitem[{{Gould}(2003)}]{gould03}
{Gould}, A. 2003, \apj, 583, 765

\bibitem[{{Grenacher} {et~al.}(1999){Grenacher}, {Jetzer}, {Str{\"a}ssle}, \&
  {de Paolis}}]{grenacher99}
{Grenacher}, L., {Jetzer}, P., {Str{\"a}ssle}, M., \& {de Paolis}, F. 1999,
  \aap, 351, 775

\bibitem[{{Griest}(1991)}]{griest91}
{Griest}, K. 1991, \apj, 366, 412

\bibitem[{{Griest} {et~al.}(1991){Griest}, {Alcock}, {Axelrod}, {Bennett},
  {Cook}, {Freeman}, {Park}, {Perlmutter}, {Peterson}, {Quinn}, {Rodgers},
  {Stubbs}, \& {The MACHO Collaboration}}]{griest91b}
{Griest}, K., {Alcock}, C., {Axelrod}, T.~S., {et~al.} 1991, \apjl, 372, L79

\bibitem[{{Gyuk}(1999)}]{gyuk99}
{Gyuk}, G. 1999, \apj, 510, 205

\bibitem[{{Hamadache} {et~al.}(2006){Hamadache}, {Le Guillou}, {Tisserand},
  {Afonso}, {Albert}, {Andersen}, {Ansari}, {Aubourg}, {Bareyre}, {Beaulieu},
  {Charlot}, {Coutures}, {Ferlet}, {Fouqu{\'e}}, {Glicenstein}, {Goldman},
  {Gould}, {Graff}, {Gros}, {Haissinski}, {de Kat}, {Lesquoy}, {Loup},
  {Magneville}, {Marquette}, {Maurice}, {Maury}, {Milsztajn}, {Moniez},
  {Palanque-Delabrouille}, {Perdereau}, {Rahal}, {Rich}, {Spiro},
  {Vidal-Madjar}, {Vigroux}, \& {Zylberajch}}]{hamadache06}
{Hamadache}, C., {Le Guillou}, L., {Tisserand}, P., {et~al.} 2006, \aap, 454,
  185

\bibitem[{{Han} \& {Gould}(1995)}]{hangould95}
{Han}, C. \& {Gould}, A. 1995, \apj, 447, 53

\bibitem[{{Han} \& {Gould}(1996)}]{hangould96}
{Han}, C. \& {Gould}, A. 1996, \apj, 467, 540

\bibitem[{{Han} \& {Gould}(2003)}]{hangould03}
{Han}, C. \& {Gould}, A. 2003, \apj, 592, 172

\bibitem[{{Holtzman} {et~al.}(1998){Holtzman}, {Watson}, {Baum}, {Grillmair},
  {Groth}, {Light}, {Lynds}, \& {O'Neil}}]{holtzman98}
{Holtzman}, J.~A., {Watson}, A.~M., {Baum}, W.~A., {et~al.} 1998, \aj, 115,
  1946

\bibitem[{{Jetzer}(1994)}]{jetzer94b}
{Jetzer}, P. 1994, \apjl, 432, L43

\bibitem[{{Kent}(1992)}]{kent92}
{Kent}, S.~M. 1992, \apj, 387, 181

\bibitem[{{Kiraga} \& {Paczynski}(1994)}]{kiraga94}
{Kiraga}, M. \& {Paczynski}, B. 1994, \apjl, 430, L101

\bibitem[{{Koz{\l}owski} {et~al.}(2006){Koz{\l}owski}, {Wo{\'z}niak}, {Mao},
  {Smith}, {Sumi}, {Vestrand}, \& {Wyrzykowski}}]{kozlowski06}
{Koz{\l}owski}, S., {Wo{\'z}niak}, P.~R., {Mao}, S., {et~al.} 2006, \mnras,
  370, 435

\bibitem[{{Kroupa}(2002)}]{kroupa02}
{Kroupa}, P. 2002, Science, 295, 82

\bibitem[{{Kroupa}(2007)}]{kroupa07}
{Kroupa}, P. 2007, ArXiv Astrophysics e-prints, astro-ph/0703124

\bibitem[{{Mancini} {et~al.}(2004){Mancini}, {Calchi Novati}, {Jetzer}, \&
  {Scarpetta}}]{mancini04}
{Mancini}, L., {Calchi Novati}, S., {Jetzer}, P., \& {Scarpetta}, G. 2004,
  \aap, 427, 61

\bibitem[{{Minchev} {et~al.}(2007){Minchev}, {Nordhaus}, \&
  {Quillen}}]{minchev07}
{Minchev}, I., {Nordhaus}, J., \& {Quillen}, A.~C. 2007, \apjl, 664, L31

\bibitem[{Paczy\'nski(1986)}]{pacz86}
Paczy\'nski, B. 1986, \apj, 304, 1

\bibitem[{{Paczynski}(1991)}]{pacz91}
{Paczynski}, B. 1991, \apjl, 371, L63

\bibitem[{{Paczynski} {et~al.}(1994){Paczynski}, {Stanek}, {Udalski},
  {Szymanski}, {Kaluzny}, {Kubiak}, {Mateo}, \& {Krzeminski}}]{pacz94}
{Paczynski}, B., {Stanek}, K.~Z., {Udalski}, A., {et~al.} 1994, \apjl, 435,
  L113

\bibitem[{{Peale}(1998)}]{peale98}
{Peale}, S.~J. 1998, \apj, 509, 177

\bibitem[{{Picaud} \& {Robin}(2004)}]{picaud04}
{Picaud}, S. \& {Robin}, A.~C. 2004, \aap, 428, 891

\bibitem[{{Popowski}(2001)}]{popowski01}
{Popowski}, P. 2001, in ASP Conf. Ser., Vol. 239, Microlensing 2000: A New Era
  of Microlensing Astrophysics, ed. J.~W. {Menzies} \& P.~D. {Sackett} (San
  Francisco: ASP), 244

\bibitem[{{Popowski} {et~al.}(2005){Popowski}, {Griest}, {Thomas}, {Cook},
  {Bennett}, {Becker}, {Alves}, {Minniti}, {Drake}, {Alcock}, {Allsman},
  {Axelrod}, {Freeman}, {Geha}, {Lehner}, {Marshall}, {Nelson}, {Peterson},
  {Quinn}, {Stubbs}, {Sutherland}, {Vandehei}, \& {Welch}}]{popowski05}
{Popowski}, P., {Griest}, K., {Thomas}, C.~L., {et~al.} 2005, \apj, 631, 879

\bibitem[{{Rattenbury} {et~al.}(2007{\natexlab{a}}){Rattenbury}, {Mao},
  {Debattista}, {Sumi}, {Gerhard}, \& {de Lorenzi}}]{rattenbury07b}
{Rattenbury}, N.~J., {Mao}, S., {Debattista}, V.~P., {et~al.}
  2007{\natexlab{a}}, \mnras, 378, 1165

\bibitem[{{Rattenbury} {et~al.}(2007{\natexlab{b}}){Rattenbury}, {Mao}, {Sumi},
  \& {Smith}}]{rattenbury07a}
{Rattenbury}, N.~J., {Mao}, S., {Sumi}, T., \& {Smith}, M.~C.
  2007{\natexlab{b}}, \mnras, 378, 1064

\bibitem[{{Rich} {et~al.}(2007){Rich}, {Reitzel}, {Howard}, \& {Zhao}}]{rich07}
{Rich}, R.~M., {Reitzel}, D.~B., {Howard}, C.~D., \& {Zhao}, H. 2007, \apjl,
  658, L29

\bibitem[{{Riffeser} {et~al.}(2006){Riffeser}, {Fliri}, {Seitz}, \&
  {Bender}}]{arno06}
{Riffeser}, A., {Fliri}, J., {Seitz}, S., \& {Bender}, R. 2006, \apjs, 163, 225

\bibitem[{{Roulet} \& {Mollerach}(1997)}]{roulet97}
{Roulet}, E. \& {Mollerach}, S. 1997, \physrep, 279, 67

\bibitem[{{Sevenster} {et~al.}(1999){Sevenster}, {Saha}, {Valls-Gabaud}, \&
  {Fux}}]{sevenster99}
{Sevenster}, M., {Saha}, P., {Valls-Gabaud}, D., \& {Fux}, R. 1999, \mnras,
  307, 584

\bibitem[{{Smith} {et~al.}(2007){Smith}, {Wo{\'z}niak}, {Mao}, \&
  {Sumi}}]{smith07}
{Smith}, M.~C., {Wo{\'z}niak}, P., {Mao}, S., \& {Sumi}, T. 2007, \mnras, 380,
  805

\bibitem[{{Stanek} {et~al.}(1997){Stanek}, {Udalski}, {Szymanski}, {Kaluzny},
  {Kubiak}, {Mateo}, \& {Krzeminski}}]{stanek97}
{Stanek}, K.~Z., {Udalski}, A., {Szymanski}, M., {et~al.} 1997, \apj, 477, 163

\bibitem[{{Sumi} {et~al.}(2006){Sumi}, {Wo{\'z}niak}, {Udalski},
  {Szyma{\'n}ski}, {Kubiak}, {Pietrzy{\'n}ski}, {Soszy{\'n}ski},
  {{\.Z}ebru{\'n}}, {Szewczyk}, {Wyrzykowski}, \& {Paczy{\'n}ski}}]{sumi06}
{Sumi}, T., {Wo{\'z}niak}, P.~R., {Udalski}, A., {et~al.} 2006, \apj, 636, 240

\bibitem[{{Tisserand} {et~al.}(2007){Tisserand}, {Le Guillou}, {Afonso},
  {Albert}, {Andersen}, {Ansari}, {Aubourg}, {Bareyre}, {Beaulieu}, {Charlot},
  {Coutures}, {Ferlet}, {Fouqu{\'e}}, {Glicenstein}, {Goldman}, {Gould},
  {Graff}, {Gros}, {Haissinski}, {Hamadache}, {de Kat}, {Lasserre}, {Lesquoy},
  {Loup}, {Magneville}, {Marquette}, {Maurice}, {Maury}, {Milsztajn}, {Moniez},
  {Palanque-Delabrouille}, {Perdereau}, {Rahal}, {Rich}, {Spiro},
  {Vidal-Madjar}, {Vigroux}, {Zylberajch}, \& {The EROS-2
  Collaboration}}]{eros07}
{Tisserand}, P., {Le Guillou}, L., {Afonso}, C., {et~al.} 2007, \aap, 469, 387

\bibitem[{{Vallenari} {et~al.}(2006){Vallenari}, {Pasetto}, {Bertelli},
  {Chiosi}, {Spagna}, \& {Lattanzi}}]{vallenari06}
{Vallenari}, A., {Pasetto}, S., {Bertelli}, G., {et~al.} 2006, \aap, 451, 125

\bibitem[{{Wood}(2007)}]{wood07}
{Wood}, A. 2007, \mnras, 380, 901

\bibitem[{{Wood} \& {Mao}(2005)}]{woodmao05}
{Wood}, A. \& {Mao}, S. 2005, \mnras, 362, 945

\bibitem[{{Zhao} \& {Mao}(1996)}]{zhao96b}
{Zhao}, H. \& {Mao}, S. 1996, \mnras, 283, 1197

\bibitem[{{Zhao} {et~al.}(1996){Zhao}, {Rich}, \& {Spergel}}]{zhao96a}
{Zhao}, H., {Rich}, R.~M., \& {Spergel}, D.~N. 1996, \mnras, 282, 175

\bibitem[{{Zhao} {et~al.}(1995){Zhao}, {Spergel}, \& {Rich}}]{zhao95}
{Zhao}, H., {Spergel}, D.~N., \& {Rich}, R.~M. 1995, \apjl, 440, L13

\bibitem[{{Zoccali} {et~al.}(2000){Zoccali}, {Cassisi}, {Frogel}, {Gould},
  {Ortolani}, {Renzini}, {Rich}, \& {Stephens}}]{zoccali00}
{Zoccali}, M., {Cassisi}, S., {Frogel}, J.~A., {et~al.} 2000, \apj, 530, 418

\end{thebibliography}

\end{document}